\documentclass[10pt,a4paper,oneside,onecolumn,titlepage,final,openany]{article}

\usepackage{amsmath}
\usepackage{booktabs,lscape}
\usepackage{multirow}
\usepackage{amsfonts}
\usepackage{amssymb,wasysym}
\usepackage{graphicx,titlesec}
\usepackage{abstract}
\usepackage{authblk}
\usepackage{physics}
\usepackage{gensymb}
\usepackage{rotating}
\usepackage{wrapfig}
\usepackage[margin=2.5cm,a4paper]{geometry}
\usepackage[sort]{cite}
\usepackage{setspace}
\usepackage{parskip}
\usepackage{microtype}
\usepackage{tensor}
\usepackage[nomarkers]{endfloat}
\usepackage[capitalise]{cleveref}
\usepackage[separate-uncertainty=true,multi-part-units=single]{siunitx}
\usepackage{newfloat}
\usepackage{url}
\usepackage[version=4]{mhchem} 
\usepackage{chemmacros}
\usepackage{xspace}
\DeclareFloatingEnvironment[name={Supporting Video},listname={Supporting Video Captions}]{suppfigure}
\DeclareDelayedFloat{suppfigure}{Supporting Videos}

\usepackage{marginnote}

\usepackage[backgroundcolor=cyan,disable]{todonotes}
\usepackage[final]{changes}
\usepackage{changebar}
\DeclareMathOperator*{\argmin}{\arg\!\min}

\DeclareDelayedFloatFlavor{sidewaysfigure}{figure}
\DeclareDelayedFloatFlavor{sidewaystable}{table}
\usepackage{fancyhdr}

\newcommand{\jjmLOF}[3]{\added{Tab.~#1 --- #2 \dotfill #3}\\}

\title{Rapid, B1-insensitive, dual-band quasi-adiabatic saturation transfer with optimal control for complete quantification of myocardial ATP flux}
\author[1,2,3]{Jack J. Miller\textdagger}
\author[3,4]{Ladislav Valkovič\textdagger}
\author[2]{Matthew Kerr}
\author[2]{Kerstin N. Timm}
\author[3]{William Watson}
\author[2,3]{Justin Y. C. Lau}
\author[2,3]{Andrew Tyler}
\author[3,5]{Christopher Rodgers}
\author[3,6]{Paul A. Bottomley}
\author[2]{Lisa C. Heather\textdaggerdbl}
\author[2,3]{Damian J. Tyler\textdaggerdbl}

\affil[1]{Department of Physics, University of Oxford, UK}
\affil[2]{Department of Physiology, Anatomy and Genetics, University of Oxford, UK}
\affil[3]{Oxford Centre for Clinical Magnetic Resonance Research, John Radcliffe Hospital, Headington, Oxford, UK}
\affil[4]{Department of Imaging Methods, Institute of Measurement Science, Slovak Academy of Sciences, SVK}
\affil[5]{Wolfson Brain Imaging Centre, University of Cambridge, UK}
\affil[6]{Division of MR Research, Department of Radiology and Radiological Science, The Johns Hopkins University School of Medicine, Baltimore, Maryland, USA}

\date{\vspace{1 em}\flushleft Running title: Rapid, B1-insensitive Saturation Transfer with Optimal Control\\\vspace{1em}Word Count: 5307. \\\vspace{1em} Article Type: Full Paper\\\vspace{1em}Corresponding author:\\ Dr. Jack J. Miller, DPhil.\\Department of Physiology, Anatomy \& Genetics\\Sherrington Building\\University of Oxford\\Parks Road\\Oxford\\OX1 3PT\\Tel:+44 1865 282161\\Fax +44 1865 282272 
\\ Email: jack.miller@physics.org\\\vspace{1em}
\emph{\centering\qquad\qquad\qquad\qquad Submitted to Magnetic Resonance in Medicine 16-Jul-2020.}
}

\usepackage{tikz}
\usetikzlibrary{matrix,calc,shapes,arrows}
\tikzset{
  treenode/.style = {shape=rectangle, rounded corners,
                     draw, anchor=center, font=\normalsize,
                     minimum height=1cm,minimum width=3cm, align=center,
                     top color=white, bottom color=white,
                     inner sep=1ex},
  decision/.style = {treenode, diamond, inner sep=0pt},
  root/.style     = {treenode, font=\Large, bottom color=red!30},
  env/.style      = {treenode, font=\normalsize},
  finish/.style   = {root, bottom color=green!40},
  dummy/.style    = {circle,draw}
}

\begin{document}

\maketitle

\begin{abstract}
\doublespacing
\emph{Purpose:} Phosphorus saturation-transfer experiments can quantify metabolic fluxes non-invasively. Typically, the forward flux through the creatine-kinase reaction is investigated by observing the decrease in phosphocreatine (PCr) after saturation of $\gamma$-ATP. The quantification of total ATP utilisation is currently under-explored, as it requires simultaneous saturation of inorganic phosphate (\added{P\textsubscript{i}}) and PCr. This is challenging, as currently available saturation pulses reduce the already-low $\gamma$-ATP signal present.\\
 \\[2em]
 \emph{Methods:} Using a hybrid optimal-control and Shinnar-Le-Roux method, a quasi-adiabatic RF pulse was designed for the dual-saturation of PCr and \added{P\textsubscript{i}} to enable determination of total ATP utilisation. The pulses were evaluated in Bloch equation simulations, compared with a conventional hard-cosine DANTE saturation sequence, before being applied to perfused rat hearts at 11.7 Tesla.\\ 
 \\[2em]
  \emph{Results:} The quasi-adiabatic pulse was insensitive to a $>2.5$-fold variation in $B_1$, producing equivalent saturation with a 53\% reduction in delivered pulse power and a 33-fold reduction in spillover at the minimum effective $B_1$. This enabled the complete quantification of the synthesis and degradation fluxes for ATP in 30-45 minutes in the perfused rat heart. While the net synthesis flux (\added{$4.24\pm0.8$} mM/s, SEM) was not significantly different from degradation flux (\added{$6.88\pm2$} mM/s, \added{$p=0.06$}) and both measures are consistent with prior work, nonlinear error analysis highlights uncertainties in the Pi-to-ATP measurement that may explain \added{a trend suggesting a} possible imbalance.\\
  \\[2em]
  \emph{Conclusion:} This work demonstrates a novel quasi-adiabatic dual-saturation RF pulse with significantly improved performance that can be used to measure ATP turnover in the heart in vivo. 
  \\[2em]
\end{abstract}

\doublespacing
\pagestyle{fancy}
\section*{Introduction}

The heart is a metabolic omnivore with the greatest energy requirement, per gram of tissue, of any organ in the body, in order to power contraction and maintain cellular membrane potentials. This requirement is met by sophisticated metabolic machinery, culminating in a high rate of adenosine triphosphate (ATP) hydrolysis and regeneration that is mediated through the ``energy buffer'' and subcellular transportation system that shuttles phosphocreatine (PCr), a moiety that forms the primary energy reserve of the heart\cite{Beer2002,Schar2010}, through the enzyme creatine kinase (CK) to regenerate ATP. This is represented by the coupled reactions:

\begin{equation}
\label{eqn:OverviewScheme}
\text{PCr} +\text{ADP} \underset{k_r}{\stackrel{k_f}{\rightleftharpoons}}  \text{ATP}  \underset{k_f'}{\stackrel{k_r'}{\rightleftharpoons}}  \text{ADP} + \text{Pi}, 
\end{equation}

where $k_f$ and $k_r$ represent the pseudo first-order forward and reverse reaction rates through the CK reaction, and $k_f'$ and $k_r'$ represent the effective reaction rates for the conversion of ATP to adenosine diphosphate (ADP) and inorganic phosphate (Pi).  Together these coupled reactions determine how much ATP is synthesi\added{z}ed and utili\added{z}ed in the heart. This reaction scheme \eqref{eqn:OverviewScheme} has been extensively studied in the heart using Phosphorus Magnetic Resonance Spectroscopy (\textsuperscript{31}P-MRS) saturation transfer approaches. Such methods permit the quantification of the rate constants for the CK reaction itself and the net rate of ATP hydrolysis and synthesis as a whole.\cite{Spencer1988} These rates are typically determined by saturating or inverting spins of one/or more exchange partner/s, e.g., $\gamma$-ATP, for a period of time, and observing the subsequent depletion in signal of the other exchange partner, e.g., PCr. If metabolite concentrations are known or measured, the total metabolic flux through any given part of the system can be calculated.

\deleted{The assessment of the rate constants of ATP synthesis and degradation have a long history in \textsuperscript{31}P spectroscopy.}\added{However, w}hile the use of saturation transfer experiments can provide quantitative measurements of $k_f$, $k_f'$ and $k_r$ and $k_r'$, it is the forward CK flux reaction $k_f$ that has  \deleted{seen most attention}\added{been extensively studied} in the heart, in part owing to the poor SNR of cardiac \textsuperscript{31}P-MRS wherein PCr, having the highest SNR, provides the most accessible readout in the saturation transfer experiment in clinical settings.  \added{This remains highly biologically relevant because} metabolic dysregulation plays a key role in common heart diseases,\cite{Neubauer2007a}and \added{as a consequence} the forward CK rate constant $k_f$ has \added{additionally} been associated with cardiac metabolic health, with reductions in $k_f$ reported in conditions ranging from obesity\added{,}\cite{J.2017,Rayner2020} \added{and diabetes,\cite{Bashir2015}} to heart failure\cite{Weiss2005,Gupta2011}  and myocardial infarction.\cite{Bottomley2009} Thus far, this is the only rate constant currently routinely measurable in the human heart in vivo. The direct quantification of both sets of forward rate constants (\cref{eqn:OverviewScheme}) is challenging in clinical settings, due to the inability to reliably quantify the myocardial \added{P\textsubscript{i}} signal in the presence of overlapping signals from diphosphoglycerate compounds originating in the blood pool for the forward saturation transfer experiment; and the typically low signal-to-noise ratio (SNR) of \added{P\textsubscript{i}} achievable in healthy tissue in clinically feasible scan times. \added{ We therefore wish to design a scheme to permit  the role of the total reverse synthetic flux, $(k_r+k_r')$ to be revealed in humans and animals, in order to reveal further biological insight.}\todo{R2.1}

The reverse experiment is much more complex, as it requires the simultaneous saturation of PCr and \added{P\textsubscript{i}} for varying durations\added{, and the determination of the relative decrease in $\gamma$-ATP that results}. The determination of the total forward and reverse flux of ATP are obtained by fitting the acquired signals to solutions of the Bloch-McConnell equations,\cite{Spencer1988,Bashir2014}. \deleted{Therefore, this experiment is rarely if ever performed in humans.} Whilst the SNR of $\gamma$-ATP is typically worse than that of PCr, the dual-saturation experiment has the potential to report on \textit{the} biologically-key energetic process: the net turnover of the whole cellular ATP pool, beyond that involved in the CK flux reaction alone. 


Additionally, $k_f$ and the forward CK reaction is  technically easier to perform than the $k_f'$/$k_r'$ experiment requiring only narrow single-band saturation pulses with a sharp stop-band. Narrow bandwidth pulses can be made longer and tailored to better minimise Gibbs ringing than the larger-bandwidth shorter pulses that are required to saturate both PCr and \added{P\textsubscript{i}} simultaneously. Furthermore, the latter are typically associated with higher peak $B_1$ fields,  increased power requirements for the non-proton RF amplifier and higher rates of RF specific absorption in the body (i.e. higher SAR).

Even so, the length of the saturation pulses used in the conventional CK flux experiment is finite and their amplitude and centre frequency difficult to adjust. It is common to have incomplete or excessive saturation and to accidentally spillover into neighbouring resonances partially saturating them. Traditionally a control experiment is performed, wherein the saturation frequency is shifted by the same frequency offset to the other side of the resonance of interest. While this typically leads to a first-order correction for unwanted loss, replacing $M_{0}^\text{PCR}$ with $\mu M_{0}^\text{PCR}$,\cite{Bashir2014} a powerful near off-resonance irradiation may cause other nonlinear phase loss effects analogous to the Bloch-Siegert effect that are not corrected for.\cite{Kingsley2000} Attempting to design ``perfect'' saturation pulses that would avoid the need for control experiments is therefore desirable for both single-band and dual-band saturation transfer experiments. Furthermore, at high fields in the complex electrodynamic environment of the heart and in clinical research employing \textsuperscript{31}P-MRS surface coils, $B_1$ is not uniform. It is therefore desirable to design adiabaticity into the RF pulse to provide tolerance to $B_1$ inhomogeneity to aid its application. 


In this work, we demonstrate a hybrid optimal-control RF pulse design scheme, with alterations to the alpha and beta polynomials after Shinnar-Le Roux (SLR) transformation that optimise the passband and stop-band homogeneity, while achieving low sensitivity to  nonuniformity in the RF excitation field, $B_1$. The novelty of this design approach is that it combines the accuracy of computationally-challenging optimal-control methods which excel at constrained problems, with the speed and analytic underpinning of the SLR approach.  Specifically, we present a highly uniform and temporally long dual-band saturation pulse that simultaneously saturates PCr and Pi, and eliminates the need to obtain control data within each transfer experiment. Whilst other adiabatic saturation pulses have been proposed in the context of \textsuperscript{31}P-MRS for applications including the CK-flux measurement (e.g. \cite{Bashir2014}), we believe that this approach is the first multi-frequency selective quasi-adiabatic pulse designed \added{explicitly for this experiment}. \todo{R2.1}\added{Although previous multi-band frequency-selective RF excitation methods and sequences have been proposed in the context of proton imaging,\cite{Grissom2009} or spectroscopy,\cite{Posse1995} hyperpolarized \textsuperscript{13}C imaging,\cite{Miller2015a,Lau2011} for ameliorating susceptibility artefacts \cite{Miller2017a,Yip2009}, their utility for spectroscopic acquisitions is typically limited by the required frequency specificity of the pulse and  its excitation sidebands. Furthermore, spectral-spatial pulses to date have predominantly been optimised for excitation, with the SLR method used to create sub-pulses in a manually imposed envelope designed explicitly \emph{a priori}, without quantitative consideration of their overall degree of adiabaticity. It is for these reasons that other schemes, such as BASING,\cite{Star-Lack1997} have previously been used in a spectroscopic context to selectively not excite undesired resonances of interest, such as water. We note that this approach is not necessarily straightforward to integrate into conventional saturation-transfer experiments.}

We demonstrate the new pulse \added{design scheme} in a saturation transfer experiment on the Langendorff retrograde perfused rat heart at \SI{11.7}{\tesla}, to measure the energetic status of the heart as a whole. This set-up is of particular value in drug studies on disease models, where the workload of the heart, oxygenation status, and pressure afterload can be independently controlled rapidly, repeatably, and with or without pharmacological alteration.\added{ It additionally does not contain blood, and thus intracellular phosphate is more readily observable.} 


\section*{Theory}
\subsection*{Pulse design}

The extreme requirements of uniform saturation, low off-resonance spillover, and a relative degree of insensitivity to $B_1$ pose a set of unique challenges to the design of multi-frequency saturation RF pulses. In comparison to low flip-angle RF pulse design, the problem of designing efficient saturation pulses is complicated by the fact that the RF is not purely the Fourier transform of the desired excitation profile. A number of methods have been proposed to function in this regime. Most famous is the SLR approach,\cite{Shinnar1989} which relies on an analytic transformation of the RF excitation into a domain which permits phrasing of the design problem as matrix algebra, which can be solved with recursive and computationally efficient filter design algorithms such as Parks-McClellan.\cite{McClellan2005}

The SLR algorithm reduces the problem of pulse design to that of finding two polynomials, $A(z)$ and $B(z)$ where $z= e^{-i\omega}$, which may be reversibly transformed to find an RF pulse\cite{Schulte2004}. It is also possible to transform between the coefficients of the $A$ and $B$ polynomials and their equivalent frequency representation via a $z$-transformation, which is related to the discrete Fourier transformation. The latter, in turn can be related to frequency-domain Bloch equation simulations via a transformation that exploits the Cayley-Klein representation of rotations.\cite{Schulte2004} Given the difficulty of inverting the Bloch equation directly, the SLR procedure operates through these two other domains directly, as schemati\added{z}ed in scheme \eqref{eqn:SLROverview}: 

\begin{equation}
\label{eqn:SLROverview}
\begin{tikzpicture}[line width=0.5mm]
  \matrix (chart)
    [
      matrix of nodes,
      column sep      = 11em,
      row sep         = 4em,
      column 1/.style = {nodes={env}},
      column 2/.style = {nodes={env}}
    ]
    {
      RF Pulse $u(t)$               &  $A(z)$, $B(z)$ Coefficients      \\
      Frequency Domain $M_{xy}(\omega)$        &   $A(\omega)$, $B(\omega)$ Frequency Domain       \\
    };
  \draw[latex-]
    (chart-2-1) edge node [left] {Bloch Equations} (chart-1-1);

  \draw[-left to]
    (chart-1-1) edge node [above] {SLR Transform} (chart-1-2);
  \draw[-left to]
    (chart-1-2) edge node [below] {Inverse SLR Transform} (chart-1-1);
  \draw[-left to]
    (chart-1-2) edge node [right] {$z$-Transform} (chart-2-2);
  \draw[-left to] 
   (chart-2-2) edge node [left] {$z^{-1}$-Transform} (chart-1-2);
  \draw[-latex]
    (chart-2-2) edge node [above] {Cayley-Klein} (chart-2-1);
  \draw[-latex]
    (chart-2-1) edge node [below] {$\alpha$, $\beta$} (chart-2-2);
\end{tikzpicture}
\end{equation}

Although SLR neglects relaxation terms and, conventionally, considerations relating to $B_1$ homogeneity, it has been widely adopted and remains a ``gold standard'' for RF pulse design. It has been shown to be possible to expand the SLR design framework to generate adiabatic pulses, that is, pulses that are insensitive to a range of $B_1$ values above a certain threshold. This is desirable, both for the heart at ultrahigh fields and for in vivo studies employing surface coils.
 
Briefly,\cite{Balchandani2010} the process  functions by manually imposing a quadratic phase in the $B(\omega)$ frequency domain, modulating the linear phase generated by a filter designed with the specified single-band pass/stop-band requirements. The $A(\omega)$ polynomial is designed under a least-phase constraint, to yield a minimum-energy (least SAR) pulse with a quadratic phase variation. Not all quadratic phase RF pulses are adiabatic, but those with a resulting frequency ramp that is slow enough to satisfy the adiabatic condition will be so.  It remains uncertain how to directly specify multiple frequencies for excitation or saturation within this framework, as it requires root-flipping procedures to minimise the peak amplitude of a fixed duration pulse, or equivalently, reducing its duration with a fixed peak amplitude. This process involves replacing the selected roots of its $A(z)$ and/or $B(z)$ polynomials with the inverses of their complex conjugates. The best pattern of flipped roots produces the most uniform distribution of RF energy in time, imitating a quadratic phase pulse.\cite{Sharma2016} Owing to its non-linear nature, the process of root flipping is not directly compatible with an imposition of a complex phase in the $B(\omega)$ domain. 

An alternative framework for designing RF pulses in MR based on optimal-control considers the Bloch equations without exchange, and minimi\added{z}es an integrated metric for the difference between their current and the desired solution. Typically this is performed with a cost term based on the square of the RF pulse amplitude incorporated as an SAR penalty. Under these conditions, the design problem falls into the analytical framework of linear quadratic control. Assuming one spatial degree of freedom -- a slice selection gradient along a slice axis $z$ -- the Bloch equations in the rotating frame can be written as 

\begin{align}
	\frac{\partial}{\partial t} M(t,\, z) &= \mathbf{A}(u(t),\, z) M(t,\, z) + b,\quad t>0 \\ 
	\nonumber\text{given } M(0, z) &= M_0
\end{align}

where $u(t) = \bigl(u_x(t),\, u_y(t)\bigr)$ describes the normali\added{z}ed, non-dimensional RF pulse and the lineari\added{z}ed Bloch rotation matrix $A$ and longitudinal relaxation vector $b$ are given by 

\begin{align}
\label{eqn:OCBloch}
	\mathbf{A}(u,\,z) &= \begin{pmatrix}
		-1/T_2 & \gamma G_z(t)z & \gamma B_{1}^\text{max} u_y(t) \\ 
		-\gamma G_z(t)z & -1/T_2 & \gamma B_{1}^\text{max} u_x(t) \\ 
		-\gamma B_{1}^\text{max}  u_y(t) & -\gamma B_{1}^\text{max} u_x(t) & -1/T_1 
	\end{pmatrix}; \quad b = \begin{pmatrix}
		0 \\ 0 \\ M_0 / T_1
	\end{pmatrix},
\end{align}

with $G_z(t)$ an applied slice-selection gradient.\cite{Aigner2016}

Within the framework of optimal-control, we seek an RF pulse $u(t)$ of a pre-specified duration $T_u$ that minimises a chosen cost function after it is played (at $t>T_u$ for time $T=T_u+\epsilon$ where $\epsilon \ll 1$) to generate a numerically-determined solution $M(T,\,z)$ as compared to a desired predefined solution, $M_d (z)$, within a finite computational domain ($z\in [-a,\,a]$). Including the quadratic cost function to accommodate the practical limitations on RF amplifier power and SAR limitations, the one-dimensional pulse design problem is 

\begin{equation}
\label{eqn:OCCost}
	\argmin_{u(t)} J(u(t))= \frac{1}{2} \int_{-a}^{a} \bigl\lVert M(T,\,z) - M_d(z)\bigr\rVert^{2}_{2} \,\dd z + \frac{\lambda}{2} \int_{0}^{T_u} \bigl \lVert u(t) \bigr \rVert_{2}^{2} \,\dd t. 
\end{equation}

Here $\lambda$ effectively acts as a regularisation term that relates the competing goals of pulse fidelity and SAR reduction. This minimisation problem is then typically approached via numerical methods. However, optimal-control algorithms remain far more computationally expensive than the matrix approach used by SLR transformation, and convergence is slow with most na\"{i}ve  gradient-descent or quasi-Gauss Newton approaches. 

Whilst \cref{eqn:OCCost} is modifiable to problems beyond one-dimension, the extension of this method into $\geq2$ dimensions becomes challenging in practice. \cite{Maximov2015} The already-slow convergence can become unacceptable with increasing dimensionality, for example, by modifying \cref{eqn:OCBloch} to include off-resonance effects or chemical exchange, or \cref{eqn:OCCost} to include information about $B_1$ variation. Whilst examples of successful applications exist\cite{Vinding2013a}, multidimensional optimal-control methods generally present as highly  computationally expensive minimisation problems that have limited their application in the design of adiabatic RF pulses that require trade‐offs between frequency selectivity, adiabaticity and pulse power.\cite{Rosenfeld1996} 

For the one-dimensional RF pulse design problem, Aigner et al.\cite{Aigner2016} recently showed that the analytic calculation of a second-order Hessian matrix of partial derivatives acting at a location in a direction can permit the development of a globally-convergent trust-region conjugate-gradient Newton method with quadratic convergence, to yield simultaneous multislice imaging pulses with excellent slice profiles. We note that with ideal gradient waveforms, the analytical treatment of simultaneous multislice excitation is directly analogous to that of the multi-band saturation needed for PCr and \added{P\textsubscript{i}} saturation in the present application. Furthermore, an ability to excite PCr and \added{P\textsubscript{i}} in antiphase analogous to a CAIPIRINHA-based simultaneous multislice excitation,\cite{Breuer2006}, could provide a simple alternate phase-cycling scheme to improve saturation efficiency. 

\section*{Methods} 

\subsection*{Pulse design}

This work combines the SLR and optimal-control approaches described above to provide dual-band saturation with minimal ripple for simultaneous PCr/\added{P\textsubscript{i}} saturation. The new pulses are given a degree of adiabaticity by the imposition of a quadratic phase in the $A$, $B$ polynomial domain. The general approach can be summari\added{z}ed as 

\begin{enumerate}
	\item Highly optimi\added{z}ed optimal-control pulses are developed subject to hardware constraints, using the analytically enhanced optimal-control framework \cite{Aigner2016}. The initial condition is either an appropriate SNEEZE RF pulse \cite{Kupce1995,Nuzillard1994} for single-band saturation, or a hard-cosine pulse for dual-band saturation. Appropriate $T_1$ and $T_2$ values are included explicitly in this framework for each metabolite, as the optimal-control framework explicitly includes relaxation in pulse design. The result is a non-adiabatic pulse $u_0(t)$. 
	\item Obtain the resulting $A$ and $B$ polynomials for each pulse, that is $(A(z),\,B(z))=\mathcal{SLR}(u_0(t))$
	\item Obtain the discrete Fourier transform of the coefficients $b$ of the $B$ polynomial, which would be equal to the linear phase profile of the pulse traditionally generated via filter-design methods in the conventional SLR approach, say $\mathcal{F}(b)=F_\text{filter}(\omega)$. 
	\item\label{step:imposition} Impose a quadratic phase in the frequency domain: $F_\text{filter}(\omega) \mapsto F_\text{filter}(\omega) \times e^{i k \omega^2}$ where $k$ is a constant that determines the rate of quadratic phase cycling across, described in detail elsewhere. \cite{Balchandani2010}
	\item Inverse the process to obtain an updated set of $B$ polynomial coefficients: $b'=\mathcal{F}^{-1}\left(F_\text{filter}(\omega) \times e^{i k \omega^2}\right)$
	\item Obtain the optimi\added{z}ed waveform with a degree of adiabaticity imposed, $u(t)=\mathcal{SLR}^{-1}(A,B')$. 
\end{enumerate}

This approach is computationally quick, and produces a far simpler optimisation problem than using a full optimal-control framework for the design of adiabatic pulses with multiple excitation bands. The imposition of quadratic phase on the $B$ polynomial does not affect the spectral bandwidth(s) of the pulse, but increases its overall duration and decreases the effective peak $B_1$ value. 

\begin{sidewaysfigure*}
	\begin{center}
		\includegraphics[width=0.7\linewidth]{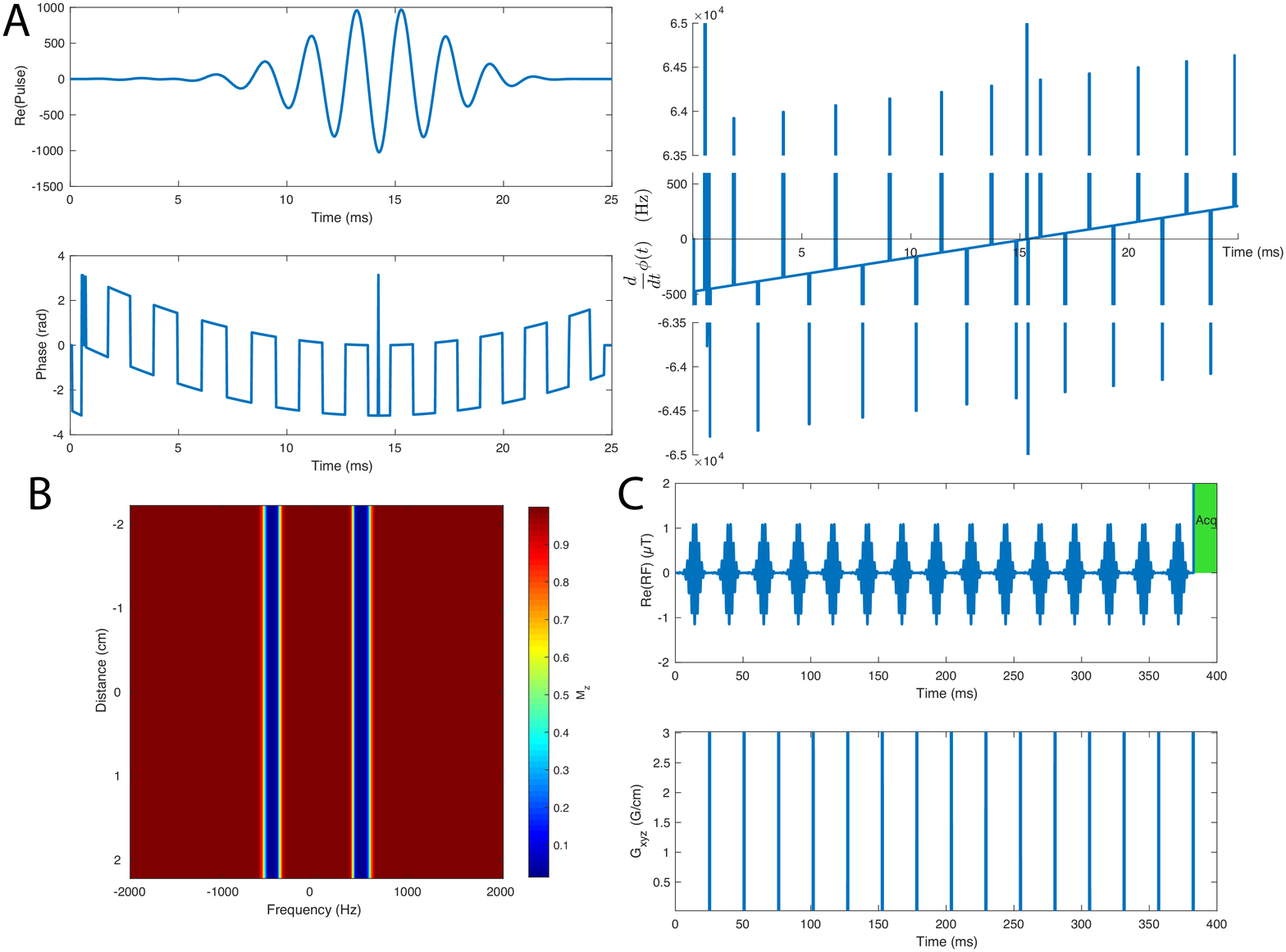}
		\caption{The designed quasi-adiabatic optimal-control pulse, \added{with both the real part of the pulse and its phase shown (\textbf{A}, top, bottom)} together with the rate of change of phase (right). This show\deleted{ing}\added{s} a clear linear frequency ramp, \added{with interleaved offsets, indicative of the adiabatic nature of the pulse. Also shown is}  \deleted{together with} its excitation profile \added{as a function of space and frequency} (\textbf{B}) \added{ expressed as the value of $M_z$ after the pulse, showing no spectral or spatial aliasing. The pulse can then be } \deleted{and integration}\added{integrated} into a saturation chain followed by a hard readout \added{(top) with gradient crushers after each excitation (bottom) as shown in } (\textbf{C})\added{ although we note that other saturation-transfer sequences or approaches could use the pulse in a `drop-in' fashion}. \todo{R2M1}}
		\label{fig:OCPulse}
	\end{center}
\end{sidewaysfigure*}

We used the above protocol to design a novel quasi-adiabatic dual-band saturation RF pulse for the \textsuperscript{31}P saturation transfer experiment\added{ at \SI{202}{MHz}}. This was a dual-band excitation with a \SI{950}{Hz} gap between saturation bands, a \SI{150}{Hz} FWHM of each pole. The pulse was \SI{25}{ms} in length compri\added{z}ed of 2500 points, and $k=4.1\times 10^{-6}$, with a nominal $B_1$ of $\SI{1.3}{\micro\tesla}$. The optimal-control algorithm was initiated with a hard-cosine pulse and a target magnetisation consisting of two slabs filtered with a \SI{2.5}{Hz} transition width convolved with a Gaussian kernel. The optimal-control methods were otherwise as described previously,\cite{Aigner2016} with a maximum relative passband excitation, that is, $\frac{|M_{xy}|}{M_0}$ at \SI{0}{Hz}, of $10^{-7}$. The simulation was allowed to converge before the manual imposition of quadratic phase, which took about 4 hours on a MacBookPro15,3 laptop computer (2.9 GHz Intel Core i9 processor). The resulting pulse is shown in \Cref{fig:OCPulse}\textbf{A}, together with its spatial and spectral response after a \SI{90}{\degree} pulse with subsequent \added{gradient} crusher (\Cref{fig:OCPulse}\textbf{B}) and implementation into a saturate/crush/excite sequence (\Cref{fig:OCPulse}\textbf{C}).

Following design, the immunity of the pulse to $B_1$ variation was quantified via Bloch equation simulations, and the saturation efficiency compared to conventional DANTE-based trains\cite{Clark1991} with hard-cosine pulses to produce dual-band excitation at the requisite frequencies. Specifically, a \SI{19}{ms} hard-cosine pulse with a \SI{500}{\micro\second} nutation frequency was chosen to provide an appropriate control, in keeping with previous work.\cite{Spencer1988} In order to produce saturation, as with the new dual-band pulses, this was repeated numerous times back-to-back in order to produce the DANTE chain. The total duration of saturation required is determined by the $T_1$ of the molecule in question and typically $3-5\times T_1$ is necessary for adequate saturation. Therefore, for subsequent relative power calculations determined from the integral of $B_{1}^{2}$, the total saturation duration was kept constant, and $B_1$ set to the minimum $B_1$ \added{needed} to produce effective saturation.
\subsection*{Experimental methods}
All experiments were performed on a vertical-bore \SI{11.7}{\tesla} spectrometer (Magnex Scientific / Varian DDR2) at a constant $\SI{37\pm0.1}{\degree C}$, with temperature maintained by blowing heated air from a Bruker BVT3000 variable-temperature NMR unit onto the back surface of the \SI{20}{mm} dual-tuned \textsuperscript{1}H/\textsuperscript{31}P volume coil and the bottom of the sample tube. This was performed using a second-hand Bruker BVT3000 variable-temperature NMR unit and an open-source re-implementation of its control software running on a Raspberry Pi written for this purpose (and released to the community by the author at \url{https://github.com/NeutralKaon/BVTserialInterfacer}).  Temperature was monitored and maintained within specimens using a separate platinum \SI{100}{\ohm} `PT100' 4-wire resistive thermal device, and using PID (proportional/integral/derivative) control through the heater. 

The new saturation pulse was initially tested on a \SI{100}{mM} phenyl-phosphoric acid (PPA) phantom. The PPA has a single peak $\sim10$ times larger than the PCr peak seen in vivo, and saturation efficiency was assessed by iteratively offsetting the transmitter voltage and frequency in 41 steps between $\pm\SI{2}{kHz}$.

For experiments on the perfused heart, male Wistar rats ($\sim\SI{250}{g}$) were sacrificed via i.p. pentobarbitone anaesthetic overdose (\SI{0.7}{ml} of \SI{200}{mg/ml}).  \deleted{and h}\added{H}earts \added{were} rapidly exci\added{z}ed and arrested in ice-cold Krebs-Henseleit buffer, containing \SI{1.2}{mM} KH$_2$PO$_4$ to prevent the slow loss of intracellular phosphate. The heart was then weighed, and rapidly cannulated via the aorta prior to Langendorff perfusion with Krebs-\deleted{Henseliet}\added{Henseleit}\todo{R1M3} buffer at \SI{37.2}{\degree C} and \SI{100}{mmHg}, delivered in a warmed umbilical. A balloon containing \SI{100}{mM} phenyl-phosphoric acid was introduced into the left ventricle (LV), and inflated to achieve an end-diastolic pressure of 3-\SI{5}{mmHg}. The heart was then placed in the warmed NMR tube and homeothermic MR coil within the magnet. All animal investigations conformed to Home Office Guidance on the Operation of the Animals (Scientific Procedures) Act of 1986 (ASPA), to institutional guidelines, and were separately approved by the University of Oxford Animal Ethics Review Committee. All compounds were obtained from Sigma Aldrich (Gillingham, Dorset, UK).

\subsection*{MR Protocol} 
The measurement protocol for assessing total myocardial energetics was compri\added{z}ed of: (i) the acquisition of localisers; (ii) shimming; (iii) a \textsuperscript{31}P spectrum based Ernst-angle frequency adjustment; (iv) acquisition of a 5-minute fully-relaxed \textsuperscript{31}P spectrum; (v) single-band and dual-band saturation-transfer experiments; and (vi)  acquisition of three one-minute fully-relaxed spectra to obtain the volume of the internal PPA phantom to use as a concentration reference (see below) and for the fully-relaxed metabolite signals at the end of the perfusion experiment. The entire protocol took 30-45 minutes, including the preparation and insertion of the perfused heart into the MR system. 

The saturation pulse was developed as a ``module'' that preceded a simple hard-pulse-acquire spectroscopic readout. This was applied (protocol step v) under fully-relaxed conditions (TR \SI{10}{s}, 16 averages, \SI{90}{\degree} flip angle, \SI{10}{kHz} bandwidth, 2048 complex points) to enable absolute quantification based on the internal PPA phantom and volume, and the weight of the heart. We note that the presence of the phantom inside the LV ensures that the phantom signal is detected with substantially the same sensitivity as that of the myocardium. Trains of either single-band or dual-band saturation pulses of total duration $[0,\,0.15,\,0.275,\,0.575,\,2.25,\,4.575]$ \si{s}, were applied, followed by all-axes gradient crusher pulses, the hard excitation pulse, and FID readout. After acquisition of fully-relaxed, and selectively saturated spectra, absolute quantification (protocol step vi) was performed by adding two known volumes of PPA (typically $\sim\SI{50}{\micro\litre}$) to the internal PPA phantom and obtaining two separate fully-relaxed spectra. This process accounts for variability in heart sizes, and hence LV balloon volumes since the balloon is inflated to a diastolic pressure that ensures retrograde perfusion via the coronary sinus and arteries during diastole.

\subsection*{Data Processing}
Spectra were fitted in the time domain via the OXSA implementation of the AMARES algorithm \cite{Purvis2017} given  prior  knowledge of the expected location of the peaks of phenyl-phosphoric acid, intracellular and extracellular phosphate, phosphocreatine, the two phosphodiesters glycerophosphocholine (GPC) and glycerophosphoethanolamine (GPE), \added{the } nicotinamide adenine dinucleotide \added{pool} (NAD/NAD(P)H; modelled as a single peak), and ATP, including $J$-coupling. A linear correction factor was determined from the fully-relaxed spectra with different PPA volumes in protocol step (vi) to obtain the volume of the \SI{100}{mM} PPA present during the saturation transfer experiments (step v). Absolute metabolite concentrations were calculated using an intracellular volume fraction for healthy myocardium of 52\% and tissue specific gravity of $\sim\SI{1.05}{g\per ml}$.\cite{Tyler2010}

Saturation transfer data were fitted to integrated solutions of the Bloch-McConnell equations via a bounded nonlinear least-squares method, in the form 
\begin{align}
Y(t)&=M_0(1-(k \cdot \tau(1-e^{-\frac{t}{\tau}}))) \\ 
\text{where } \tau &= \frac{1}{T_1}+k;  \nonumber
\end{align}

and where $k=k_f$ if $Y$ is the the amplitude of PCr, $k_f'$ of Pi, or $(k_r+k_r')$ if it is $\gamma$-ATP following dual saturation. For this, a global optimisation algorithm was used with the linear constraint that $\tau-k\geq0$ and all parameters $\geq0$.\cite{Conn1991} All values reported are mean $\pm$ SEM.

\section*{Results} 
\begin{figure}
\begin{center}
	\includegraphics[width=\linewidth]{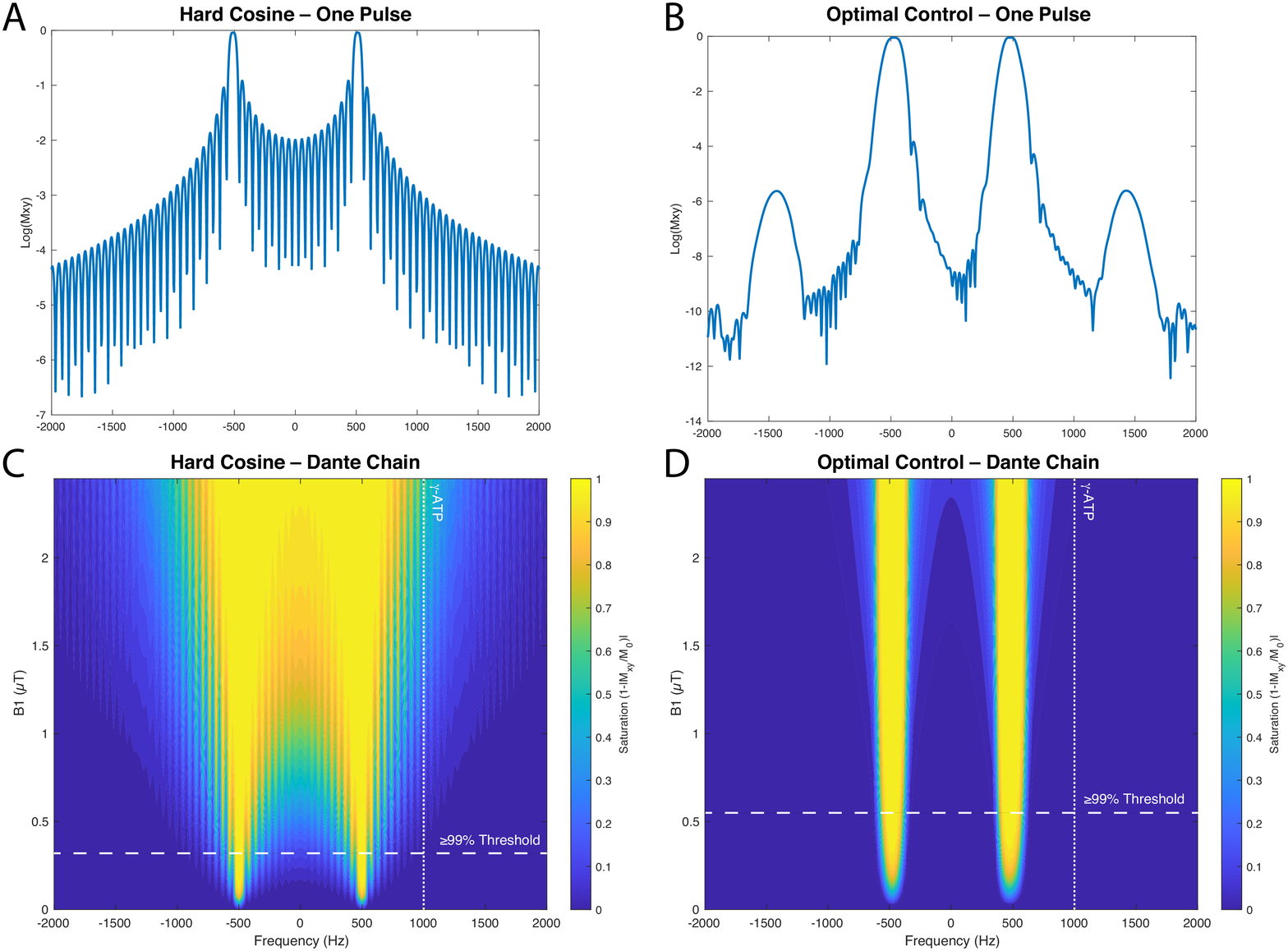}
	\caption{A comparison of the designed quasi-adiabatic optimal control pulse when compared to the conventional hard-cosine approach. After a single excitation is played (\textbf{A}, \textbf{B}) the designed pulse exhibits a significantly decreased degree of erroneous excitation outside of the designed spectral passband as shown by the base-ten logarithm of transverse magnetisation. As a consequence, when integrated into a DANTE saturation chain, the novel pulse is effectively immune to plausible variation in $B_1$, in stark comparison to a hard-cosine pulse (\textbf{C}, \textbf{D}), which functions as designed within a comparatively small window of acceptable $B_1$ given a desire to minimise unwanted saturation of $\gamma$-ATP.}
	\label{fig:OCSimulation}
\end{center}	
\end{figure}

Bloch equation simulations revealed that the designed dual-band quasi-adiabatic pulse demonstrates a far greater immunity to $B_1$ variation, with the optimi\added{z}ed dual-band pulse featuring a dramatic increase in immunity to variation compared to the conventional hard-cosine pulses  when used either for \SI{90}{\degree} excitation (\Cref{fig:OCSimulation}\textbf{A}, \textbf{B}) or in a DANTE chain (\textbf{C}, \textbf{D}). As pulse power increases further, the behaviour of the new pulse remains benign, with minor variations in the total effective saturation occurring within the designed saturation bandwidth, rather than between and beyond them. This reflects on the optimisation process which provides both very flat passbands and  immunity to $B_1$ variation. The pulse frequency modulation ($\partial\phi/\partial t$) has three linear but interleaved frequency ramps with constant offsets, characteristic of adiabatic pulses.\cite{Balchandani2010}

Compared to an equivalent hard-cosine DANTE chain saturating both PCr and \added{P\textsubscript{i}} with $>99$\% saturation efficiency on both resonances, the pulse required a 70\% higher peak $B_1$ (\SI{0.32}{\micro\tesla} vs \SI{0.55}{\micro\tesla}) but possessed a $B_{1\text{ rms}}$ that was 50\% lower: \SI{0.0516}{\micro\tesla} for the hard-cosine vs \SI{0.0275}{\micro\tesla}. This corresponded to a lower integrated power deposition ($P\propto\int_{0}^{T_u} |B_1(t)|^2\,\dd t$) over the duration of the longest DANTE saturation chain considered, i.e. \SI{4.58}{s}: the deposited power of the new pulse is just 53.4\% of that for the hard-cosine pulse. Similarly, at the frequency offset corresponding to $\gamma$-ATP (i.e. $\sim\SI{-1000}{Hz}$ compared to the mid-point of \added{P\textsubscript{i}} and PCr), the designed pulse featured a relative excitation of $\sim 10^{-9}$ in comparison to $10^{-3}$ for a conventional hard-cosine pulse. This improvement in selectivity, by six orders of magnitude, translates into a dramatic reduction in the degree of spill-over following saturation by a DANTE chain. At the frequency  of $\gamma$-ATP, a 10\% erroneous saturation occurs if the hard-cosine pulse is played at or above a peak $B_1$ of \SI{0.63}{\micro\tesla}, whereas the designed novel pulse does not reach that point until peak $B_1$ exceeds \SI{4.2}{\micro\tesla}. This therefore gives it a greatly expanded range in which $B_1$ variation does not significantly affect the $\gamma$-ATP signal; the saturation remains $\leq 1\%$ until peak $B_1>\SI{1.5}{\micro\tesla}$, yielding a $\geq 2.5$-fold effective immunity to $B_1$ variation. When played at the minimum effective $B_1$, spillover at $\gamma$-ATP is therefore $\sim 0.03\%$ compared to 1\% for the hard-cosine pulse, 33-fold lower. 

\begin{figure}
	\begin{center}
		\includegraphics[width=0.5\linewidth]{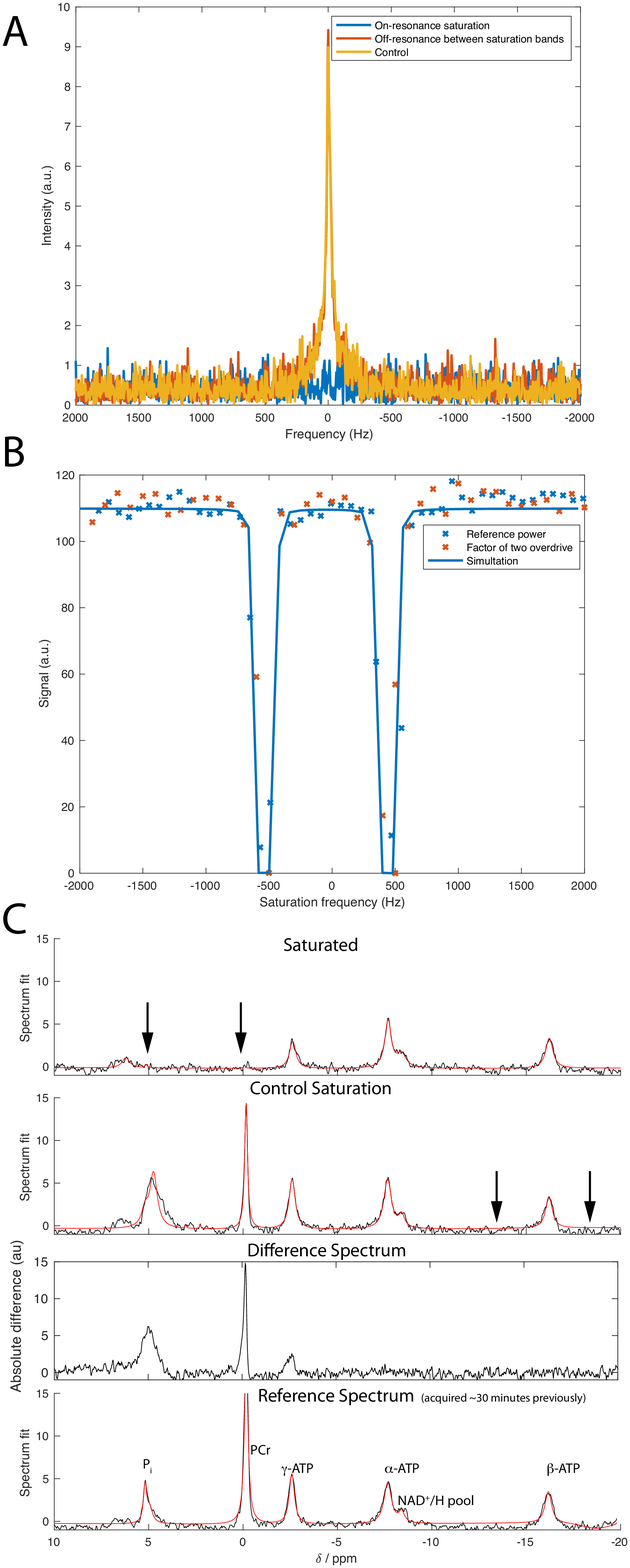}
		\caption{\textbf{A}: Example phantom saturation data on resonance (i.e. fully-saturating the phantom peak), off-resonance between the saturation bands (i.e. at \SI{0}{Hz}), and a separately acquired control acquisition with no dual-band saturation pulse played. \textbf{B}: The resulting behaviour as a function of saturation power is identical over an experimental factor of two variation in $B_1$\added{. A 50 Hz offset in the frequency axis was introduced between experiments to enable the visualisation of these points}\todo{R1M4}. \textbf{C}:\todo{R1.3} In the perfused heart, the performance of the saturation pulse is the same, with complete saturation of intracellular inorganic phosphate and PCr, and no detectable alteration in the $\beta$-ATP peak. \added{A \SI{20}{Hz} exponential apodization has been applied. The location of the saturation pulse is indicated via black arrows, and the difference between played and control saturation schemes is shown below. A hard pulse/acquire scheme was used for the `reference spectrum' example, which was temporally acquired before the other acquisitions. For this reason, the exact PCr/ATP ratio is slightly different as the heart acclimatized (For further peak assignments, c.f. \cref{fig:InVivoData})} \todo{R1M5, R1.3}}
		\label{fig:PhantomData}
	\end{center}
\end{figure}

The predicted frequency-domain response was verified with the PPA phantom (\Cref{fig:PhantomData}\textbf{A}). A control experiment \added{undertaken to verify both the quasi-adiabatic nature of the pulse and explicitly map its saturation profile as a function of frequency was}\todo{R1M4} performed by shifting the frequency of the saturation pulse \added{at two different power levels}\todo{R1M4}. The results (\Cref{fig:PhantomData}\textbf{B}) show excellent agreement between the predicted \added{and} experimental pulse behaviour.

Single-band and dual-band saturation experiments performed on $n=\deleted{4}\added{5}$ na\"ive  Wistar rat hearts show that that no measurable saturation correction is needed when the pulse is applied off-resonance, as illustrated in \Cref{fig:PhantomData}\textbf{C}. 

\begin{sidewaysfigure}
\begin{center}
	\includegraphics[width=\linewidth]{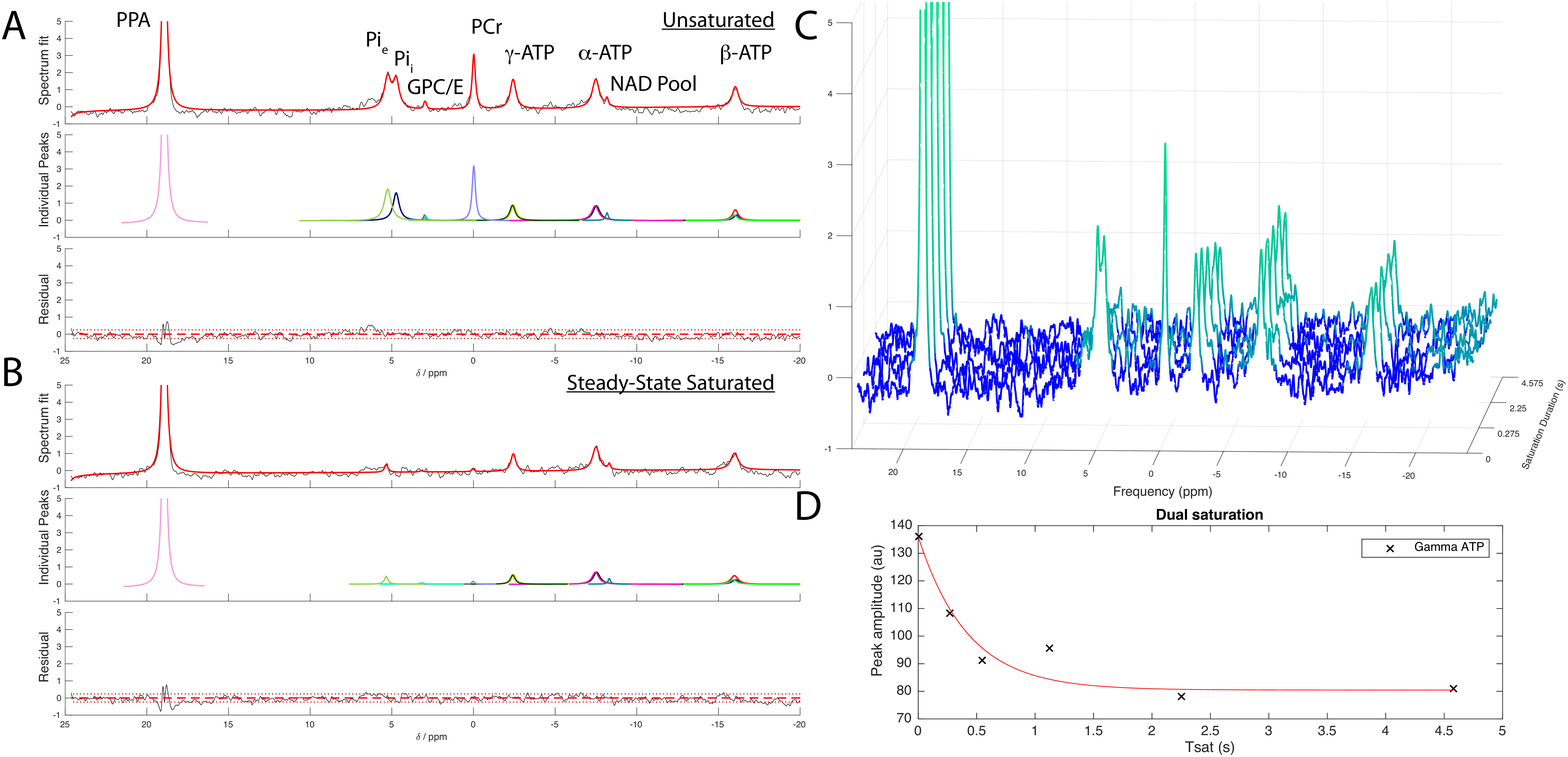}
	\caption{\todo{R2M2}The integration of the designed dual-band quasi-adiabatic pulse in the ex vivo perfused heart. \textbf{A, B}: Simultaneous and complete saturation of phosphocreatine and the intracellular and extracellular \added{inorganic phosphate}\todo{R1M6} was observed, with concomitant decreases due to saturation transfer in ATP, shown with both individual AMARES fitting components and fit residuals. Saturated peak intensities were either numerically found to be zero, or the standard deviation of the noise floor, shown by a dotted red line. \added{A \SI{20}{Hz} apodization has been applied.} \textbf{C, D}: The time-evolution of the dual-band saturation transfer experiment shows, as expected, approximately exponential behaviour. \added{(Abbreviations: PPA -- phenylphosphoric acid; P$_\text{ie}$, P$_\text{ii}$-- extra/intracellular inorganic phosphate; GPC/E -- glycerophospho-choline/-ethanolamine; PCr -- phosphocreatine; ATP -- adenosine triphosphate.)}\todo{R1M5}}  
	\label{fig:InVivoData}
\end{center}	
\end{sidewaysfigure}

\added{As illustrated in \cref{fig:InVivoData}(\textbf{A, B}), t}he\todo{R2M2} new saturation transfer protocol also permitted the absolute quantification of all major phosphorus compounds of interest \added{ with no measurable spill over. As expected, the time behaviour of metabolites appeared to decay exponentially with the duration of saturation (\cref{fig:InVivoData}\textbf{C, D}). When combined with absolute quantification of metabolites, provided here via an internal phantom, it is thus possible to quantify the concentration of \textsuperscript{31}P-containing moieties absolutely in the heart, and the developed saturation pulse permits the quantification of reaction rates. } Multiplying the reaction rates by the corresponding concentrations, yielded the total ATP synthesis and degradation fluxes for the heart in approximately 30-45 minutes, a duration compatible with perfused heart experiments. \added{Spectra are shown with a \SI{20}{Hz} apodization in the form of exponential line-broadening. Whilst it is the case that \textsuperscript{31}P MRS is inherently insensitive, the ability of the AMARES algorithm to quantify both peak amplitudes and the Cram\'{e}r-Rao Lower Bound of uncertainty on those amplitudes permits a quantitative analysis of the limitations of the technique, as detailed in the Online Supporting Information. In our work,}\todo{R1.1} the mean ATP concentration was \added{$\SI{7.31\pm 0.5}{mM}$} and the mean PCr concentration was \added{$\SI{11.6\pm0.7}{mM}$} (\Cref{tab:allData}). The synthetic rate constants $k_f$ and $k_f'$ were \added{$\SI{0.27\pm0.05}{\per\second}$} and \added{$\SI{0.21\pm0.06}{\per\second}$}, respectively with a degradative rate of  \added{$(k_r+k_r') = \SI{1.05\pm0.19}{\per\second}$}. The corresponding synthesis and degradation fluxes for ATP were thus \added{$\SI{4.24\pm0.82}{mM\per\second}$} and \added{$\SI{6.88\pm1.82}{mM\per\second}$}\added{ ($p=0.06$ via paired $t$-test; $p=0.23$ via unpaired unequal variance $t$-test; $p=0.4$ via Wilxocon test)}. Heart function\added{, as quantified via developed pressure,} did not decrease significantly throughout the duration of perfusion.

\begin{sidewaystable}[]
\centering
\begin{tabular}{@{}cccccccccccccc@{}}
\toprule
\multirow{2}{*}{Heart}  & [$\gamma$-ATP] & [PCr] & PCr/$\gamma$-ATP & [Pi] & $k_f$            & $k_f'$           & $(k_r+k_r')$     & Synth. Flux        & Deg. Flux          & Heart Weight & Heart Rate & LVDP & RPP               \\
                        & mM             & mM    &                  & mM   & \si{\per\second} & \si{\per\second} & \si{\per\second} & \si{mM\per\second} & \si{mM\per\second} & g            & bpm        & mmHg & \si{mmHg\per min} \\ \midrule
1                       & 7.56           & 10.44 & 1.38             & 6.93 & 0.43             & 0.42             & 1.77             & 7.35               & 13.39              & 1.92         & 133        & 133  & 17600                  \\
2                       & 7.02           & 10.95 & 1.56             & 7.01 & 0.25             & 0.21             & 0.99             & 4.23               & 6.93               & 1.26         & 216        & 114  & 24600                  \\
3                       & 6.90           & 10.15 & 1.47             & 4.38 & 0.28             & 0.11             & 0.92             & 3.30               & 6.34               & 1.16         & 204        & 131  & 26800                  \\
4                       & 9.12           & 12.59 & 1.38             & 5.21 & 0.26             & 0.09             & 0.60             & 3.72               & 5.52               & 1.10         & 228        & 106  & 24100                  \\ 
\added{5}    & \added{5.96}   &\added{13.78} & \added{2.31}& \added{2.98} & \added{0.14}             & \added{0.22}             & \added{0.96} & \added{2.58}               & \added{2.22}               & \added{1.20}         & \added{122}       & \added{187} & \added{22800} \\

\midrule
Mean & \added{7.31} &  \added{11.58} & \added{1.62} &  \added{5.30} & \added{0.27} & \added{0.21} & \added{1.05} &  \added{4.24} & \added{6.88} & \added{1.33}  & \added{180.60} & \added{134}  & \added{23180}  \\
SEM & \added{0.52} & \added{0.69} & \added{0.18} & \added{0.77} & \added{0.05} & \added{0.06} & \added{0.19} & \added{0.82} & \added{1.82} & \added{0.15} & \added{22.08} & \added{14} & \added{1537} \\

\bottomrule
\end{tabular}
\caption{Summary of derived $^{31}$P quantities of interest in a cohort of $n=5$ naive Wistar rat hearts as determined through the described technique. (LVDP: left-ventricular developed pressure; RPP: rate-pressure product).\todo{R1.1, R1.2, R2.2} }
\label{tab:allData}
\end{sidewaystable}

\section*{Discussion}

Here we have demonstrated the development of a quasi-adiabatic dual-band saturation pulse that is able to efficiently and simultaneously saturate both \added{P\textsubscript{i}} and PCr with a $\geq 2.5$-fold immunity to $B_1$ at ultrahigh field, and hence permit the determination of $(k_r+k_r')$. The constraint on  $B_{1}^{2}$ applied in the optimal-control scheme used for designing pulses, reduces \added{the} SAR \added{required to achieve}\deleted{for} practical pulses that can be used in animal models and \added{could be }adapted for human studies at \SI{3}{T} and \SI{7}{T} systems  within the same \added{design} framework. Compared to an optimi\added{z}ed single-band saturation pulse \added{that} broadly saturat\deleted{ing}\added{es} both PCr and Pi, the dual-band pulse is significantly longer with a lower $B_{1\,\text{rms}}$, \added{has} sharper transition bands, and \added{offers} minimi\added{z}ed off-resonance spill-over irradiation. The use of dual-band saturation  permitted the rapid  quantification of the total ATP degradation flux from a fully-relaxed \textsuperscript{31}P spectrum, and one additional saturation transfer experiment. The increased temporal resolution potentially afforded by the omission of the control saturation data allows for dynamic experiments and extends the viability of the perfused heart model, wherein cardiac performance slowly deteriorates after a $\sim \SI{2}{hour}$  window. We note that the designed quasi-adiabatic multiple-band saturation or excitation pulses could be incorporated into more widely used saturation schemes, such as FAST,\cite{Bottomley2002} TRiST,\cite{Schar2010} STReST,\cite{Clarke2019} and TWiST,\cite{Schar2015}, and find other applications beyond  \textsuperscript{31}P MRS. \added{It may also find wider applications in that require selective excitation, including}\deleted{Examples could include} simultaneous multi-slice imaging with a degree of immunity to $B_1$ variation; the selective excitation of metabolites in hyperpolari\added{z}ed metabolic imaging; \added{chemical exchange saturation transfer (CEST),} and spatial tagging of multiple arteries or planes in the arterial spin labelling experiment.\todo{R2.1}

\added{Whilst powerful, non-invasive, and scientifically important, we note that \textsuperscript{31}P MRS is both inherently a low SNR technique, and one subject to a number of potentially confounding factors. Whilst the low SNR is both quantifiable and can be analytically understood (leading, for example, to concomitant increases in statistical power\cite{Miller2017}) the biochemical confounding factors are harder to understand analytically. These facts are true of both localised spectroscopic read-outs of the concentration of molecules within heart tissue, and, by extension, saturation-transfer experiments. We note that the perfused heart experiment as performed here is free from  spectral contamination due to blood within the myocardium (which introduces 2,3-DPG resonances that overlap with P\textsubscript{i}), permitting the determination of intracellular P\textsubscript{i}. It is also guaranteed to be free of spectral contamination from nearby skeletal muscle and liver, such as may occur in the \emph{in vivo} situation depending upon the spatial response function of the localization sequence used. It additionally offers the potential to measure  cardiac work, and thus quantify both cardiac energy supply and utilization to support contractile function, although this was not demonstrated here. \todo{R1.1} As a model system, the perfused heart does slowly deteriorate over time, which typically manifests as slight changes in pH (leading to a reported heterogeneous distribution in true intracellular pH)\cite{Lutz2013}, phosphocreatine, and potentially alterations in both the relative proportions and total quantity of NAD-containing compounds. The latter are altered by changes in redox potential, and these peaks spectrally overlap partially with ATP. Such slight changes and alterations are known to be most prevalent at the start of the perfusion experiment as the heart slowly acclimatizes to its new environment.\cite{Sutherland2000,Lateef2015}  Although the Langendorff heart remains a commonly-studied model system, these properties are deleterious: in addition to the apparent variability owing to the low SNR of the technique, the inherent biological variability in the preparation studied (and its further variability over time) are all confounding factors. }\todo{R1.2, R2.2}

The results of the experimental studies \added{conducted here nevertheless} are in \added{quantitative} agreement with previously-reported values of $k_f$: our \added{$\SI{0.27\pm0.05}{\per\second}$} compares to $\SI{0.35\pm0.06}{\per\second}$ in the in vivo healthy control rat heart\cite{Bashir2015} and $\SI{0.32\pm0.05}{\per\second}$ in the healthy human heart\cite{Bashir2014}. For concentrations, we have  \added{$[\text{PCr}] =\SI{11.58\pm0.69}{mM}$} vs. $\SI{11.5\pm1.01}{mM}$\cite{Tyler2010} and \added{$[\text{ATP}] =\SI{7.31\pm0.52}{mM}$} vs. $\SI{6.75\pm0.5}{mM}$ in  perfused rat heart. Similarly, our reverse rates of \added{$(k_r+k_r')=\SI{1.05\pm0.19}{\per\second}$} agree with a value of $\SI{1.15\pm0.19}{\per\second}$ measured by Spencer et al.~using hard cosine saturation pulses.\cite{Spencer1988} 
\added{However, it remains the case that } \deleted{Nevertheless}, the mean value of the total flux of ATP synthesis determined with conventional single-band saturation transfer experiment \deleted{is roughly a half that of} \added{trends lower than}  the degradation flux as measured by the dual-band experiment although the difference \deleted{is}\added{was} not statistically significant\deleted{, this result is troublesome. If it were real,}\added{. If confirmed,} it would suggest that the heart was in the process of dying and [ATP] \deleted{would}\added{should} not have been detectable in the three PPA calibration scans at the end of the experiment. In the steady-state, it is expected that these fluxes remain balanced \deleted{up} until the \deleted{point of}\added{rapid decline immediately prior to} death, consistent with the common observation that [ATP] is effectively constant in vivo and in situ, although a gradual decline with time is seen in Langendorff preparations.\todo{R1.2, R1.3}\added{ 
There are several approaches that could be used to ameliorate this problem. Firstly, the comparatively low SNR of the P\textsubscript{i} peak is the single greatest source of uncertainty in the total estimated flux, and thus improving its quantification is of greatest importance. This could be achieved without increasing the total scan time either by increasing $B_0$, or, given the comparatively small size of the perfused heart, the use of a cryocoil. Secondly, increasing the number of averages, and thus the total scan time would lead to greater spectral SNR, but potentially at the cost of the variability and stability of the perfused heart. Thirdly, another alternative is to increase $n$, the number of experiments performed, and increase statistical power and average out the effect of the uncertainty in measurements.}

\added{We note that, despite these inherent sources of variability that phosphorus spectroscopy has found a successful niche as a powerful research tool, where results from numerous patients or participants can be averaged. In this proof-of-principle work, we  explicitly determine the contribution to the uncertainties in ATP synthesis and degradation fluxes by analysis of the }\deleted{. It is possible that the inequality observed in mean flux may result from} nonlinear error propagation of both saturation transfer schemes. In particular, whilst $k_f$ can be determined accurately owing to the larger size of the PCr peak, the determination of $k_f'$ is much more challenging owing to the smaller size of the intracellular \added{P\textsubscript{i}} peak\added{, which can only readily be resolved in a cardiac setting in the perfused heart}. The addition of phosphate buffer to the perfusate increases the size of this peak via the slow intracellular import of phosphate making it easier to measure, but not fundamentally changing the kinetics through ATP-synthase.\added{ This slow import will, however,  make the determination of total flux mildly dependent upon the time after perfusion at which it was measured, because it is necessary to multiply the concentration of inorganic phosphate by $k_f'$ to calculate the total flux.} Consequently, although our determined $k_f'={\SI{0.21\pm0.06}{\per\second}}$ is not significantly different to that reported by Spencer et al.~of $\SI{0.37\pm0.09}{\per\second}$, the combined effect of measurement uncertainty (quantified by the Cram\'{e}r-Rao Lower Bound, CRLB, during spectral quantification) on both the rate constant and absolute concentration may combine unfavourably.

Analytically, if $\sigma_\text{PCr}$ etc denote the uncertainty on a reported value of [PCr] etc, estimated by its CRLB, with $\sigma_{k_f}$ etc for the rate constants determined by a curve-fitting regime, then a Taylor expansion for error propagation for the synthesis (Synth) and degradation (Deg.) flux shows, given

\begin{align}
\text{Synth. Flux} &= k_f\text{[PCr]} + k_f'\text{[Pi]} &\Rightarrow  	\frac{\sigma_\text{Synth.}}{\text{Synth}}  &= 
\sqrt{\left(\frac{\sigma_\text{[Pi]}}{\text{[Pi]}}\right)^2 + \left(\frac{\sigma_\text{[PCr]}}{\text{[PCr]}}\right)^2  + \left(\frac{\sigma_{k_f}}{k_f}\right)^2 +\left(\frac{\sigma_{k_f'}}{k_f'}\right)^2}\label{eqn:UncertaintySynth} \\ 
\text{Deg. Flux} &= (k_r+k_r')\text{[ATP]}&\Rightarrow  	\frac{\sigma_\text{Deg.~~}}{\text{Deg.}} &= 
\sqrt{\left( \frac{\sigma_{k_r+k_r'}}{k_r+k_r'} \right)^2 + \left( \frac{\sigma_\text{[ATP]}}{\text{[ATP]}} \right)^2 }.\label{eqn:UncertaintyDeg} 
\end{align}

Numerically, the estimated uncertainty in \cref{eqn:UncertaintySynth} is significantly larger than that in \cref{eqn:UncertaintyDeg} (approximately 0.75 vs 0.21 in our studies), reflecting the contribution of multiple sources of error. \added{For a detailed description of how this value is obtained, please see the Online Supporting Information. This error arises largely because of uncertainty in estimating the concentration of intracellular phosphate. This uncertainty itself may reflect another potential biochemical explanation for the apparent discrepancy. If reproduced by further experiments to be ``real'', it may originate because not all inorganic phosphate in the system is truly MR visible: even at ultrahigh field, phosphate bound to macromolecules may not be detectable, and the proportion bound will vary with physiological state. This variability may reflect a fundamental limitation of the application of these techniques in the setting of the perfused heart and presents a further complication to what are already challenging experiments. We note, however, that, independent of this limitation, \emph{in vivo} studies in humans or animals may be better served by computing only $(k_r+k_r')$ or $k_f$ directly. If it is assumed that the subject is physiologically stable, and that the MR-visible ATP synthesis and degradation fluxes are equal, then we would expect to see less variability in the determined rate constants of interest, and the remaining one could be inferred from flux balance. This approach would rely upon measuring $\gamma$-ATP accurately as a function of time, something that is routinely performed in human phosphorus spectroscopy, in addition to the $k_f$ conventional CK-flux experiment. The difficulty in resolving inorganic phosphate thus is neatly sidestepped in the reverse experiment: as there is no need to resolve it in order to saturate it effectively, and effective saturation could be confirmed by the simultaneous depletion of phosphocreatine, which can easily be monitored.}


\section*{Conclusion}

This work demonstrates a novel approach for \added{the design and application}\deleted{designing and an application} of a quasi-adiabatic dual-frequency saturation pulse, suitable for measuring forward and reverse metabolic reaction rates. The design employs a hybrid optimal-control and SLR transformation to achieve sharp frequency selection and a degree of $B_1$ insensitivity suitable for ultrahigh field NMR in experimental animals and the human heart at $\geq\SI{7}{\tesla}$. The resultant pulse was exceptionally flat, with no experimentally detectable ``spillover''. It was incorporated into a protocol to assess the \deleted{complete} high-energy phosphate compounds and \deleted{energetic status}\added{the ATP synthesis and degradation rates
} in the perfused rat heart at \SI{11.7}{\tesla}, providing measures in good agreement with the literature.  We hope this technology will be useful in future studies intent on quantifying the total flux of ATP turnover in the heart\added{, and in other areas of MR where quasi-adiabatic multi-frequency saturation is required.}


\section*{Acknowledgements} 

All authors would like to thank the British Heart Foundation for their generous support (refs RG/11/9/28921, FS/14/17/30634, FS/17/58/33072 and FS/15/68/32042), the University of Oxford British Heart Foundation Centre for Research Excellence (RE/13/1/30181) and the NHS National Institute for Health Research Oxford Biomedical Research Centre programme. The views expressed are those of the authors and not necessarily those of the NIHR or the Department of Health and Social Care. JJM would like to acknowledge a Postdoctoral Fellowship run in collaboration with Novo Nordisk and the University of Oxford, and thank financial support provided by St Hugh's College and Wadham College in the University of Oxford. LV and CTR are funded by a Sir Henry Dale Fellowship from the Royal Society and the Wellcome Trust (098436/Z/12/B).  LV also acknowledges the support of Slovak grant agencies VEGA (2/0003/20) and APVV (15‐0029). JYCL would like to acknowledge funding from the NIHR Oxford Biomedical Research Centre and support from the Fulford Junior Research Fellowship at Somerville College.
 AT would like to acknowledge funding from the Engineering and Physical Sciences Research Council (EPSRC) and Medical Research Council (MRC) [grant number EP/L016052/1]. PAB was supported by a Newton Abraham Visiting professorship at Oxford. 
\textdagger \textdaggerdbl These authors contributed jointly to this work \\\vspace{1em}
\section*{\added{Supporting Information}}

\added{Please see Online Supporting Information for an extended discussion of uncertainty propagation, including (Tab.~S1) summary uncertainty values in derived concentrations; and a summary of the data obtained for each individual experiment (Tab.~S2--Tab.S11).}
\subsection*{\added{List of Supporting Information Table Captions}}
\jjmLOF{S1}{\ignorespaces Summary uncertainty values in derived concentrations.\relax }{Page S5}
\jjmLOF{S2}{\ignorespaces Summary of parameters (Heart 1)\relax }{Page S6}
\jjmLOF{S3}{\ignorespaces Raw metabolite MRS peak amplitudes and CRLBs, and determined metabolite concentration and $\Delta $Concentration. (Heart 1)\relax }{Page S6}
\jjmLOF{S4}{\ignorespaces Summary of parameters (Heart 2)\relax }{Page S7}
\jjmLOF{S5}{\ignorespaces Raw metabolite MRS peak amplitudes and CRLBs, and determined metabolite concentration and $\Delta $Concentration. (Heart 2)\relax }{Page S7}
\jjmLOF{S6}{\ignorespaces Summary of parameters (Heart 3)\relax }{Page S8}
\jjmLOF{S7}{\ignorespaces Raw metabolite MRS peak amplitudes and CRLBs, and determined metabolite concentration and $\Delta $Concentration. (Heart 3)\relax }{Page S8}
\jjmLOF{S8}{\ignorespaces Summary of parameters (Heart 4)\relax }{Page S9}
\jjmLOF{S9}{\ignorespaces Raw metabolite MRS peak amplitudes and CRLBs, and determined metabolite concentration and $\Delta $Concentration. (Heart 4)\relax }{Page S9}
\jjmLOF{S10}{\ignorespaces Summary of parameters (Heart 5)\relax }{Page S10}
\jjmLOF{S11}{\ignorespaces Raw metabolite MRS peak amplitudes and CRLBs, and determined metabolite concentration and $\Delta $Concentration. (Heart 5)\relax }{Page S10}

\bibliographystyle{mrm}
\todo{R1M7}
\bibliography{Bibliography}
\end{document}